\documentclass[journal]{IEEEtran}

\usepackage{mathptmx}
\usepackage{array}
\usepackage{multirow}
\usepackage{tabularx}
\usepackage{amsmath}
\usepackage{graphicx}
\usepackage[colorlinks=true, allcolors=blue]{hyperref}
\usepackage{float}
\usepackage[normalem]{ulem}
\usepackage{cite}
\usepackage{amsfonts}
\usepackage{pifont}
\usepackage{afterpage}
\usepackage{stfloats}

\usepackage{threeparttable}

\newcommand{\superior}{\(\checkmark\)}
\newcommand{\intermediate}{\(\circ\)}
\newcommand{\inferior}{\(\times\)}
\begin{document}

\title{Wavefront Engineering for Physical Layer Security in Near-Field Wireless Communications}

\author{{\IEEEauthorblockN{Sotiris Droulias and Angeliki Alexiou,~\IEEEmembership{Member,~IEEE}} }
\thanks{The authors are with the Department of Digital Systems, University of Piraeus, Piraeus 18534, Greece (Corresponding author: Sotiris Droulias, e-mail: sdroulias@unipi.gr)}}
\makeatletter
\def\@IEEEaftertitletext{\vspace{-2.25em}}
\makeatother
\maketitle

\begin{abstract}
With wireless transmission exposing information to potential interception by unintended receivers, physical layer security (PLS) offers a promising approach to securing wireless communications.
Beamforming can mitigate this vulnerability by exploiting the spatial characteristics of wireless propagation to concentrate power toward the intended receiver while suppressing it at potential eavesdroppers. However, as wireless systems enter the radiating near field of large-aperture antenna arrays, the resulting wavefront curvature provides additional opportunities for spatial wavefront engineering to enhance PLS.
In this work, we investigate the use of different beam types for PLS, including focused, curved, abruptly autofocusing, and cosine beams, generated using uniform linear and planar arrays (ULAs/UPAs). To quantify their performance, we introduce power-based metrics that characterize the power received by the legitimate user and a potential eavesdropper. 
For all beam types we introduce focusing along the vertical UPA direction, to independently control the phase distributions along the two UPA dimensions and enhance the power discrimination between the legitimate receiver and the eavesdropper.
We also propose hollow-focused beams, in which an inactive strip is introduced in the transmitting aperture to create a protected low-power region between the transmitter and the legitimate receiver. The results demonstrate that vertical focusing dramatically enhances the PLS performance and that hollow-focused beams achieve superior PLS performance in both line-of-sight (LoS) and vicinity-of-user (VoU) scenarios.
\end{abstract}


\maketitle

\section{INTRODUCTION}
\IEEEPARstart{P}{hysical} layer security (PLS) aims to protect wireless communications against unauthorized interception by exploiting the inherent characteristics of the wireless physical layer \cite{Mukherjee2014, Hanzo2016, Schaefer2017}. Unlike conventional cryptographic approaches, PLS leverages the properties of wireless propagation to enhance the confidentiality of transmitted information. A fundamental challenge stems from the broadcast nature of wireless propagation, whereby the transmitted signal is not inherently confined to the intended receiver and may consequently be intercepted by unintended users. Thus, controlling the spatial distribution of the transmitted power is central to enhancing the security of wireless links \cite{Niu2026}. \\
\indent Beamforming addresses this vulnerability by exploiting the spatial characteristics of electromagnetic propagation to concentrate power toward the legitimate receiver while suppressing leakage toward potential eavesdroppers \cite{Khisti2010, Zurita2011, Sharma2022}. Conventional beamforming, however, is predominantly designed for far-field operation, where the wavefront is approximately planar and beam control is primarily directional. The increasing use of large-aperture arrays and higher frequencies is driving wireless systems toward the radiating near field, where wavefront curvature becomes significant. This additional spatial degree of freedom can be tailored to enable beam profiles with properties beyond conventional directional transmission, offering new opportunities to control the spatial distribution of power and enhance PLS.
\subsection{Prior works}
\noindent Recent works have investigated several unconventional beam types for near-field wireless communications, including focused beams \cite{Eldar2022, Dai2023, Droulias2024}, curved beams \cite{Guerboukha2024, Droulias2025, Uchimura2026}, abruptly autofocusing (AAF) beams \cite{Droulias2025}, and cosine beams \cite{Droulias2025_cosine}. These beams exhibit distinct propagation characteristics that can potentially be exploited for secure transmission. \\
\indent Focused beams, in particular, have recently received considerable attention, owing to their ability to concentrate electromagnetic power within a localized spatial region. The possibility to increase PLS with near-field beamforming was studied in \cite{Martínez2022,Liu2023} for downlink transmissions, in \cite{Ferreira2023} for uplink communications, and in \cite{Ferreira2024} for both uplink and downlink, also demonstrating the increase in the jamming rejection. In \cite{Zhang2024} a two-stage algorithm was developed to maximize the secrecy capacity of the user via jointly optimizing unit-modulus phase shifters and baseband digital beamformers. The work in \cite{Eldar2024} introduced analog beamfocusing techniques utilizing true-time delayers and phase shifters to mitigate beamsplit effects in wideband communications, and to significantly enhance PLS. In \cite{DrouliasPLS_2024}, the superiority of focused beams over conventional beamforming in PLS was demonstrated using explicit electromagnetic modeling. The superiority of focused beams was also demonstrated in \cite{Zhao2025} using channel modeling to derive closed-form expressions for the secrecy capacity under a power budget and the minimum power required to achieve a target secrecy rate. Near-field secure communications in a multi-user MISO system was recently studied in \cite{Zhang2026}, also proposing a low-complexity approach, tailored for beamformer design. \\
\indent Curved beams have also recently attracted attention, owing to their ability to circumvent a potential eavesdropper that attempts to tap the communication link. The concept of wavefront hopping was recently introduced in \cite{Petrov2024, Petrov2025}, whereby Airy-type beams are combined with other beam types, such as Bessel beams, to maximize PLS. The approach involves distributing different portions of a message across distinct beams, which are alternated over time to communicate with the legitimate user. As a result, an eavesdropper is unable to intercept all message segments and, consequently, cannot fully decode the transmitted information. The first experimental demonstration of PLS using Airy beams was recently reported in \cite{Stults2026}. Lastly, in \cite{DrouliasPLS_2026}, the superiority of curved beams over conventional beamforming was demonstrated for line-of-sight (LoS) eavesdropping, as well as when the eavesdropper is in the vicinity of the user (VoU). \\
%
\indent The concept of wavefront engineering for PLS has only recently begun to attract attention. Although initial studies have investigated the potential of focused and curved beams, a wide range of alternative beam configurations remains unexplored. Further investigation and systematic comparison of different beam types is therefore needed to identify the most suitable beam characteristics from a PLS perspective.

\subsection{Our contributions}
\noindent In this work, we investigate the use of near-field wavefront engineering for enhancing PLS through the generation and evaluation of different beam types using uniform linear and planar arrays (ULAs/UPAs). The main contributions of this work are summarized as follows:
\begin{itemize}
\item A systematic investigation of focused, curved, AAF, and cosine beams for physical layer security is presented. The considered beams are generated using ULAs and UPAs, and are evaluated using power-based metrics that characterize their ability to enhance the received power at the legitimate receiver while suppressing it at a potential eavesdropper in both LoS and VoU scenarios.

\item The additional spatial degree of freedom offered by UPAs is exploited to independently control the phase distributions along the two aperture dimensions. This enables the distinctive propagation characteristics of each beam type to be preserved along the horizontal array direction while introducing cylindrical focusing along the vertical array direction, thereby enhancing the spatial power discrimination between the legitimate receiver and the eavesdropper.

\item Hollow-focused beams are proposed for both ULA- and UPA- based operation, by introducing an inactive strip within the transmitting aperture. This configuration creates a low-power protected region between the transmitter and receiver while retaining the power-concentration capability of focused beams. The resulting trade-off between the protected-region width and focusing efficiency is characterized, demonstrating the superiority of hollow-focused beams in terms of PLS performance, for both LoS and VoU scenarios.

\end{itemize}
\subsection{Organization}
The remainder of this paper is organized as follows. Section \ref{sec:SystemModel} introduces the system model and physical layer security aspects for conventional beamforming. Section \ref{sec:WavefronEngineering} presents the beam types and the proposed UPA-based wavefront design. Section \ref{sec:PerformanceMetrics} defines the power-based performance metrics. Section \ref{sec:PerformanceAssessment} evaluates the considered beams in LoS and VoU eavesdropping scenarios for ULA-based operation. Section \ref{sec:FromULAStoUPAS} extends the performance to UPA-based operation with cylindrical focusing in the vertical array direction, and introduces and analyzes the hollow-focused beam configuration. Finally, Section \ref{sec:Conclusion} concludes the paper.

\section{System model and conventional beamforming}
\label{sec:SystemModel}

\begin{figure}
    \centering
    \includegraphics[width=\linewidth]{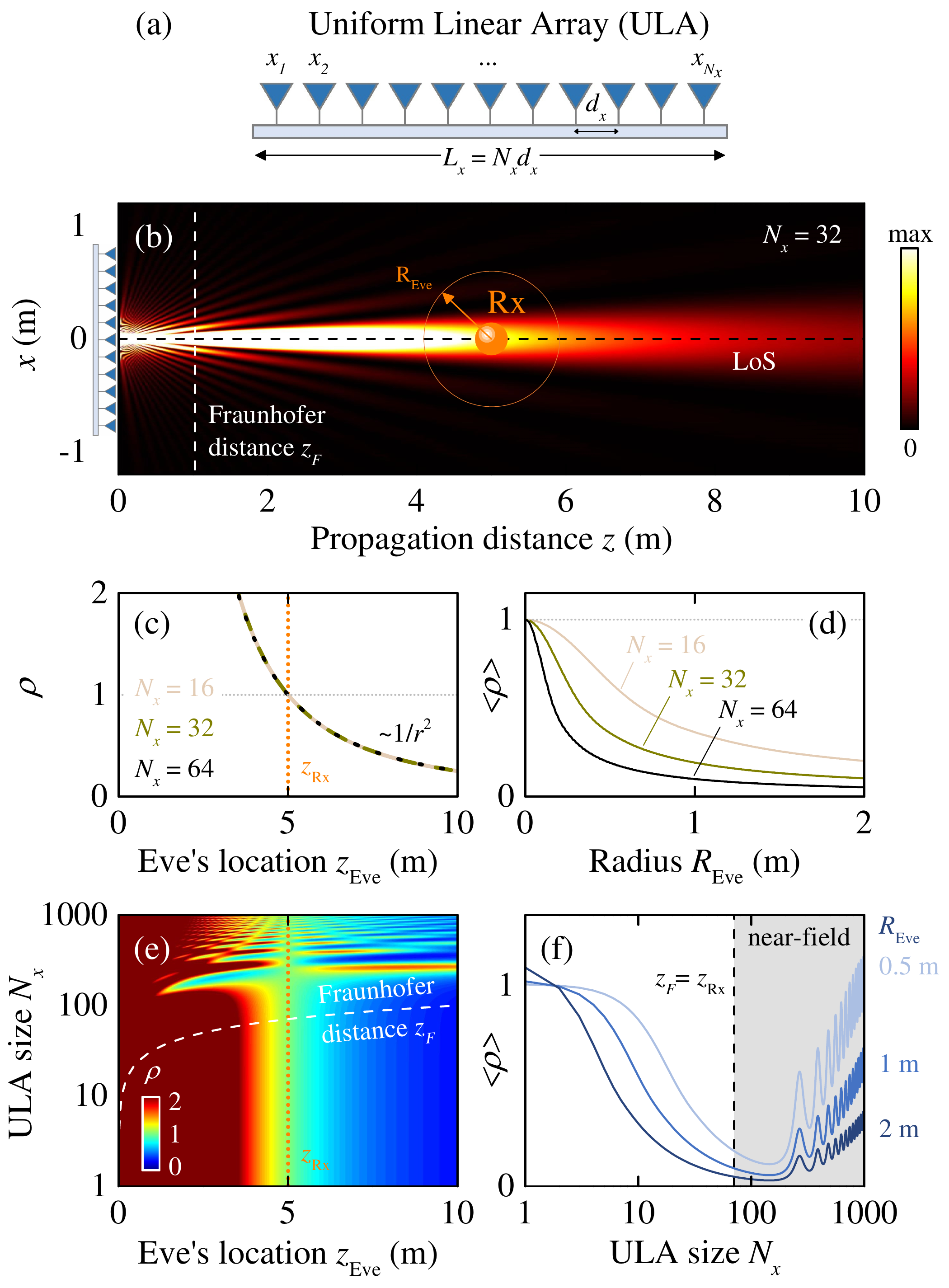}
    \caption{Conventional beamforming with ULAs. (a) ULA schematic. (b) Beamforming with a ULA of $N_x=32$ elements at 150 GHz. (c) Power ratio $\rho=P_\mathrm{Eve}/P_\mathrm{Rx}$, as a function of Eve's location along the Tx-Rx LoS, for a user located at $(x_\mathrm{Rx}, z_\mathrm{Rx})=(0\,\mathrm{m},5\,\mathrm{m})$. (d) Power ratio $\rho$, averaged within a disk of radius $R_\mathrm{Eve}$, centered at the Rx location. (e) Power ratio $\rho$ along the Tx-Rx LoS, as a function of the ULA size. (f) Averaged power ratio $\langle\rho\rangle$ for $R_\mathrm{Eve}=0.5,1,2\,\mathrm{m}$, as a function of the ULA size.}
    \label{fig:fig01}
\end{figure}

\noindent We consider a communication link scenario, in which the transmitter (Tx) is equipped with a UPA and the receiver (Rx) with a single-antenna. The UPA is located at the origin of the coordinate system and is oriented on the $xy$-plane, generating beams that propagate on the $xz$-plane toward $z>0$. The electric field of the beam at the input plane ($z=0$) has the general form
\begin{align}
    E(x,y)=A(x,y) e^{j\phi(x,y)},
    \label{Eq:EqINPUTWAVE}
\end{align}
where $A$ is the amplitude and $\phi$ the phase. The beam profile \eqref{Eq:EqINPUTWAVE} takes discrete values at the locations of the UPA elements 
\begin{align}
    x_n=\left(n-\frac{N_x+1}{2} \right)d_x, \,\,\,y_m=\left(m-\frac{N_y+1}{2} \right)d_y,
\end{align}
where $d_x$, $d_y$ is the inter-element spacing along the respective direction and $n=1,2,\dots,N_x$, $m=1,2,\dots,N_y$.
The UPA consists of $N_x\times N_y$ elements, hence its size is $L_x\times L_y=N_x d_x \times N_y d_y$. For $N_y=1$, the UPA reduces to a ULA (uniform linear array), a schematic of which is shown in Fig.\,\ref{fig:fig01}(a). In what follows, we will keep the coordinates $x,y$ as continuous variables to derive our analytical results, and in the simulations we will use the beam profile \eqref{Eq:EqINPUTWAVE} at the discrete locations $x_n, y_m$. Because beam formation is determined primarily by the input phase $\phi$ \cite{Droulias2025}, the amplitude $A$ across all UPA elements is taken as constant, $A(x,y)=A_0$, simplifying the excitation conditions in realistic systems. Also, we consider the typical inter-element spacing $d_x=d_y=\lambda/2$.\\
\indent Conventional beamforming refers to beams generated typically with beamformers. These beams are characterized by rapid intensity oscillations in the near field, and the well-known $\mathrm{sinc}$ intensity profile in the far-field, with respect to the observation angle. To steer such a beam along a steering direction $\theta_r$, the ULA phase has the well-known form
\begin{align}    
    \phi(x) = -k\sin{\theta_r}x,
    \label{Eq:EqPHASE1}
\end{align}
where $k=2\pi/\lambda$ is the free-space wavenumber and $\lambda$ the operating wavelength. One such beam with $\theta_r=0^\circ$ is shown in Fig.\,\ref{fig:fig01}(b), generated by a ULA consisting of $N_x=32$ elements ($N_y=1$) and operating at $150\,\mathrm{GHz}$. The Fraunhofer distance for this ULA is $z_F=2L_x^2/\lambda\approx1\,\mathrm{m}$, and is marked with the white dashed line. \\
\indent To simulate the propagation of the beam, we use the discrete model described in  \cite{Droulias2024_DM}, in which the $E$-field at the observation point $(x_\mathrm{obs},y_\mathrm{obs},z_\mathrm{obs})$, $E_\mathrm{obs}$, is calculated numerically as 
\begin{equation}
    E_\mathrm{obs} = \sum_{n=1}^{N_x}\sum_{m=1}^{N_y} E_{nm},
    \label{Eq:EqEr}
\end{equation}
where $E_{nm}$ is the electric field from the $nm^\mathrm{th}$ ULA element to the observation point, which is at distance $r_{nm} = \sqrt{(x_\mathrm{obs}-x_n)^2 + (y_\mathrm{obs}-y_m)^2 + z_\mathrm{obs}^2}$. The $E_{nm}$-field is given by
\begin{equation}
    E_{nm}= \sqrt{2Z_0 G_{nm} U_{nm} \frac{P_{nm}}{4\pi}} \frac{e^{-j k r_{nm}}}{r_{nm}} e^{j\phi(x_n,y_m)}, 
    \label{Eq:EqEn}    
\end{equation}
where $Z_0$ is the free-space wave impedance, $G_{nm}$ is the element gain, $U_{nm}$ the element radiation pattern, and $P_{nm}$ the power radiated per element.

%
%
%
%

\indent Let us now consider an eavesdropping scenario, in which the ULA is using this beam to communicate with a user located at $(x_\mathrm{Rx}, z_\mathrm{Rx})=(0\,\mathrm{m},5\,\mathrm{m})$. The eavesdropper (Eve) is attempting to tap the communication link between the Tx and the Rx, while moving along the Tx-Rx LoS, marked with the black dashed line in Fig.\,\ref{fig:fig01}(b), i.e. Eve is located at $(x_\mathrm{Eve}=0\,\mathrm{m}, z_\mathrm{Eve})$. We refer to this case as the \textit{LoS eavesdropping scenario}. \\
\indent The exposure of the legitimate transmission to Eve can be quantified in terms of the relative received signal power at Eve, $P_\mathrm{Eve}$, and the legitimate user, $P_\mathrm{Rx}$, i.e. the ratio
\begin{align}    
    \rho =\frac{P_\mathrm{Eve}}{P_\mathrm{Rx}},
    \label{Eq:RHO}
\end{align}
which is a measure of the success of eavesdropping. Lowering the transmit power does not generally change $\rho$ if both receivers are affected by the same power scaling; it lowers both $P_\mathrm{Eve}$ and $P_\mathrm{Rx}$. Therefore, adjusting the transmit power such that the legitimate user receives only the minimum power required for reliable detection can drive the eavesdropper's received signal below the noise floor, thereby preventing reliable signal detection and eavesdropping \cite{Goel2008, Zurita2011, Hanzo2016, Choi2023}. 
Note that, in the far-field, the beam power density decreases with distance $r$ according to the inverse-square law $\propto 1/r^2$. Therefore, the ratio $\rho$ remains high along the Tx-Rx LoS, particularly between the Tx and the Rx, where it exceeds unity, as shown in Fig.\,\ref{fig:fig01}(c). Reducing or even slightly increasing the ULA size leaves the Rx in the far-field of the Tx, therefore the power ratio remains almost unchanged. \\
%
%
%
\indent In general, the eavesdropper may also be located near the Rx without necessarily lying along the LoS path. We refer to this case as the \textit{vicinity-of-user (VoU) eavesdropping scenario}. 
Here, we evaluate the spatially averaged power ratio $\langle\rho\rangle$, assuming that Eve's location is uniformly distributed within a disk of radius $R_\mathrm{Eve}$ centered at the Rx, as illustrated in Fig.\,\ref{fig:fig01}(b). As shown in Fig.\,\ref{fig:fig01}(d), $\langle\rho\rangle$ decreases with increasing radius $R_\mathrm{Eve}$, since Eve is increasingly likely to occupy regions outside the narrow beam where little or no power is received. Increasing the ULA size further narrows the beam and consequently reduces $\langle\rho\rangle$. Does this imply that larger ULAs inherently provide more effective eavesdropping suppression? \\
\indent Increasing the ULA size produces narrower beams but simultaneously increases the Fraunhofer distance, shifting the transition to the radiating near field to larger distances. As the ULA grows, the Rx eventually enters the near field, where rapid intensity oscillations replace the conventional $1/r^2$ far-field decay, as shown in Fig.\,\ref{fig:fig01}(e). For LoS eavesdropping, this results in an oscillatory power ratio $\rho$ close to unity and, consequently, ineffective eavesdropping suppression. In the VoU case, increasing the ULA size initially reduces $\langle\rho\rangle$, as shown in Fig.\,\ref{fig:fig01}(f), but this trend reverses once the Fraunhofer distance exceeds the Rx distance. In our example, this transition occurs at $N_x\approx70$, while for $N_x=1000$, a typical array size in near-field communications, conventional beamforming provides no substantial improvement over smaller ULAs. \\
\indent These observations highlight the need for beamforming strategies capable of exploiting, rather than being limited by, near-field propagation. In the following, we demonstrate that wavefront engineering enables the spatial redistribution of beam power to maximize the power at the Rx while maintaining low power at Eve in a controllable manner.
\section{Wavefront engineering}
\label{sec:WavefronEngineering}
\subsection{Focused beams}
Focused beams concentrate the transmitted power onto a confined region in space, the so-called \textit{focal point}. By design, they converge towards the focal point and diverge beyond it, hence they can circumvent a possible eavesdropper that is located in the vicinity of the focal point. Because the rays of the focused beams are all normal to concentric spheres that converge to the focal point, beam focusing requires the formation of a converging spherical wavefront \cite{Droulias2024}. Therefore, the necessary phase is given by 
\begin{align}    
    \phi(x,y) = k \sqrt{\left(x-d_f\sin{\theta_r} \right)^2 + y^2+\left(d_f\cos{\theta_r}\right)^2},
    \label{Eq:EqPHASE2}
\end{align}
where $d_f$ is the focal distance and $\theta_r$ is the steering angle. For ULAs, the $y$-direction is suppressed and the spherical phase reduces to cylindrical phase.

\subsection{Curved beams}
\indent Curved beams propagate on bent paths and are therefore ideal for circumventing possible eavesdroppers. In this work we consider the counterpart of the well-known Airy beam, which evolves on a parabolic trajectory ($x_c,z_c$) with vertex located at ($x_0,z_0$), i.e.
\begin{align}
    x_c=x_0+\beta (z_c-z_0)^2.
    \label{Eq:EqCAUSTICparabolic}
\end{align}
\noindent The subscript $c$ stands for \textit{caustic}, which is the envelope of all the rays that form the curved beam in the geometric optics approach; it corresponds to the trajectory of the beam's main lobe in the wave approach. The caustic intersects the UPA at $x_{0C}=x_0+\beta z_0^2$, i.e. $x_{0C}$ marks the location of the trajectory at $z=0$. To maximize the UPA usage, the caustic must intersect the UPAs edge, i.e. $x_{0C}=\mathrm{sgn}(\beta) L_x/2$. The sign of the trajectory curvature parameter $\beta$ determines the direction along which the beam evolves in its far-field, that is, towards $x>0$ for $\beta>0$ and towards $x<0$ for $\beta<0$. For such a curved beam, the necessary phase is \cite{Droulias2025,DrouliasPLS_2026}
\begin{align}    
    \phi(x) = 2\beta k z_0 x + \frac{4}{3} \beta^2 k \left(z_0^2 + \frac{x_0-x}{\beta} \right)^\frac{3}{2},
    \label{Eq:EqPHASE3}
\end{align}
which is real-valued for $x < x_{0C}$, $\beta > 0$ and $x > x_{0C}$, $\beta < 0$.\\
\indent To control the extent of beam bending, it is possible to define a unique trajectory that evolves through a point $P$, located at $(x_P,z_P)$, to reach the Rx. The coordinates of $P$ must satisfy the caustic \eqref{Eq:EqCAUSTICparabolic}, i.e. $x_P =x_0 + \beta(z_P-z_0)^2$, and the same must hold for the Rx, i.e. $x_\mathrm{Rx} =x_0 + \beta(z_\mathrm{Rx}-z_0)^2$. Taking into account that the beam trajectory at $z_c=0$ crosses the $x$-axis at $x_c=x_{0C}$, i.e. $x_{0C} =x_0 + \beta(0-z_0)^2$, we end up with three equations, which are solved in terms of $x_0,z_0,\beta$, to provide the parameters of the parabolic trajectory, that is
\begin{subequations}
    \begin{align}
    \label{Eq:EqX0}
        x_0 &= \frac{\left(x_{0C}\left(z_\mathrm{Rx}-z_P \right)+\frac{x_\mathrm{Rx}z_P^2-x_Pz_\mathrm{Rx}^2}{z_\mathrm{Rx}-z_P} \right)^2-4x_{0C}x_\mathrm{Rx}z_P^2}{4\frac{z_\mathrm{Rx}z_P}{z_\mathrm{Rx}-z_P}\left(z_P\left(x_{0C}-x_\mathrm{Rx}\right)- z_\mathrm{Rx}\left(x_{0C}-x_P\right)\right)},\\ 
    \label{Eq:EqZ0}
        z_0 &= \frac{1}{2}\frac{z_P^2\left(x_{0C}-x_\mathrm{Rx}\right)- z_\mathrm{Rx}^2\left(x_{0C}-x_P\right)}{z_P\left(x_{0C}-x_\mathrm{Rx}\right)- z_\mathrm{Rx}\left(x_{0C}-x_P\right)},\\
     \label{Eq:EqB}        
        \beta &= \frac{z_P\left(x_{0C}-x_\mathrm{Rx}\right)-z_\mathrm{Rx}\left(x_{0C}-x_P\right)}{z_Pz_\mathrm{Rx}\left(z_P-z_\mathrm{Rx}\right)}.
    \end{align}
\end{subequations}
Of particular interest is the case where the point $P$ is along the Tx-Rx LoS, ensuring a protected region that extends from $P$ to the Rx. In this case, we may express the Rx coordinates in terms of the Tx-Rx distance $r_\mathrm{Rx}$, the Tx-$P$ distance $r_P$ and the LoS angle $\theta_\mathrm{LoS}$, as $(x_\mathrm{Rx}=r_\mathrm{Rx}\sin{\theta_\mathrm{LoS}},z_\mathrm{Rx}=r_\mathrm{Rx}\cos{\theta_\mathrm{LoS}})$ and $(x_P=r_P\sin{\theta_\mathrm{LoS}},z_P=r_P\cos{\theta_\mathrm{LoS}})$. In this case \eqref{Eq:EqX0}-\eqref{Eq:EqB} reduce to
\begin{subequations}
    \begin{align}
    \label{Eq:EqX01}
        x_0 &= -x_{0C}\frac{\left(r_P-r_\mathrm{Rx}\right)^2}{4 r_Pr_\mathrm{Rx}}+\sin{\theta_\mathrm{LoS}}\left(\frac{r_P+r_\mathrm{Rx}}{2} - \sin{\theta_\mathrm{LoS}} \frac{r_P r_\mathrm{Rx}}{4x_{0C}} \right),\\
    \label{Eq:EqZ01}
        z_0 &= \frac{r_P+r_\mathrm{Rx}}{2}\cos{\theta_\mathrm{LoS}}-\frac{r_P r_\mathrm{Rx}}{4x_{0C}}\sin{2\theta_\mathrm{LoS}},\\
    \label{Eq:EqB1}        
        \beta &= \frac{x_{0C}}{r_Pr_\mathrm{Rx}\cos^2{\theta_\mathrm{LoS}}}.
    \end{align}
\end{subequations}
In essence, for a certain $r_\mathrm{Rx}$, the value of $r_P$ determines the beam parameters that ensure a protected LoS region within the range $r_P$--$r_\mathrm{Rx}$. For Rx located exactly in front of the UPA, where $\theta_\mathrm{LoS}=0^\circ$ ($x_P=x_\mathrm{Rx}=0$), \eqref{Eq:EqX0}-\eqref{Eq:EqB} reduce to
\begin{subequations}
    \begin{align}
    \label{Eq:EqX02}
        x_0 &= -\frac{x_{0C} \left(z_P-z_\mathrm{Rx}\right)^2}{4 z_Pz_\mathrm{Rx}},\\
    \label{Eq:EqZ02}
        z_0 &= \frac{z_P+z_\mathrm{Rx}}{2},\\
    \label{Eq:EqB2}        
        \beta &= \frac{x_{0C}}{z_Pz_\mathrm{Rx}}.
    \end{align}
\end{subequations}
\subsection{Abruptly auto-focusing beams}
\noindent Curved beams belong to the general class of abruptly autofocusing (AAF) beams \cite{Droulias2025}, namely beams that are capable of converging and focusing at a selected target distance, while maintaining a low maximum intensity along the entire path propagated from the source. AAF has been investigated for radially symmetric beams, and is also possible for one-dimensional (1D) beams; such beams perform AAF on the $xz$-plane only, and their shape extends uniformly along the $y$-direction, thus offering larger regions of extremely low power than conventional radially symmetric beams. These beams can be excited with the combination of opposite directed curved beams, i.e. using \eqref{Eq:EqPHASE3} 
with
\begin{align}
\beta=
    \begin{cases}
        +|\beta|, & -\frac{L_x}{2}<x<0,\\
        -|\beta|, & 0<x<+\frac{L_x}{2}.
    \end{cases}    
    \label{Eq:EqPHASE4}
\end{align}

\noindent 1D AAF beams can focus behind potential blockers or possible eavesdroppers, and are therefore ideal candidates for PLS. Because they are formed by two opposite-directed curved beams, the region bounded by the two parabolic arcs forms a lenticular shape, which is symmetric about its long axis and short axis. The long axis measures $2z_0$, the distance from the UPA at which the power is concentrated, and the maximum width along the short axis measures $2x_0$, the extent of the gap between the two opposite-directed curved beams. Both caustics pass through $x=z=0$, hence $x_{0C}=0$. Applying \eqref{Eq:EqCAUSTICparabolic} for $x_\mathrm{Rx}=0$, leads to the beam parameters
\begin{subequations}
    \begin{align}
    \label{Eq:EqAAFx0}
        x_0 &= -\beta\left(\frac{z_\mathrm{Rx}}{2}\right)^2,\\ 
    \label{Eq:EqAAFz0}
        z_0 &= \frac{z_\mathrm{Rx}}{2}.
    \end{align}
\end{subequations}
Hence, the size of the long and short axis can be used as input parameters to obtain $\beta$. In practice, the power is concentrated along the $z$-axis at a focal distance that is slightly farther than $2z_0$, because the maximum of the main lobe is displaced from $x_c$ by $\delta x_m$. For the Airy beam, this displacement is given analytically \cite{Droulias2025}.

\subsection{Cosine beams}
\noindent Cosine beams belong to the general class of Bessel beams. Unlike Bessel beams, which are radially symmetric, a cosine beam is formed by the superposition of two waves traveling at opposite oblique angles on the same plane. A straightforward way to implement such beams is to introduce linear phase shifts with opposite slope around $x=0$ \cite{Droulias2025_cosine}. For steering on the $xz$-plane along the angle $\theta_r$, the necessary phase profile at the input plane is
\begin{align}
    \phi(x)=-k x \sin{\theta_r}+k\beta_x|x|,    
    \label{Eq:EqPHASE5}
\end{align}
\noindent where $\beta_x$ is a constant that controls the convergence angle of the two counter-propagating waves. This parameter can be expressed explicitly in terms of the maximum distance $z_\mathrm{max}$ at which the two waves cross each other, and the UPA width along the $x$-direction, as
\begin{align}
    \beta_x=\frac{L_x}{2z_\mathrm{max}}.
    \label{Eq:EqBETA}
\end{align}
Cosine beams offer protected regions beyond the distance $z_\mathrm{max}$, and are therefore ideal candidates for suppressing eavesdropping along the LoS, when Eve is located behind the Rx. Because cosine beams are practically invariant along the $y$-direction, similarly to 1D AAF beams, their shape on the $xz$-plane extends uniformly along the $y$-direction, thus offering larger regions of extremely low power than conventional radially symmetric Bessel beams. 
\section{Performance metrics}
\label{sec:PerformanceMetrics}
\noindent The power ratio $\rho$ provides a useful metric for evaluating PLS performance; however, it does not account for the effects of noise. This distinction is important, as a power ratio that may permit successful wiretapping in a noise-free environment could, in the presence of noise, substantially reduce or even completely prevent the possibility of successful eavesdropping. 
Therefore, in practical scenarios, a suitable metric that explicitly accounts for the SNR (signal-to-noise ratio) is the secrecy rate, which is defined as the difference between the channel capacities of the legitimate user and the eavesdropper \cite{Barros2006}, i.e.,
\begin{align}
        S = \log_2(1+\mathrm{SNR_{Rx}})-\log_2(1+\mathrm{SNR_{Eve}}),
\label{Eq:EqSR1}
\end{align}
where $\mathrm{SNR_{Eve}}=P_\mathrm{Eve}/P_N$ is the SNR at Eve, $\mathrm{SNR_{Rx}}=P_\mathrm{Rx}/P_N$ is the SNR at the Rx, and $P_N$ is the noise power. Using the definitions of $\rho$, $\mathrm{SNR_{Eve}}$ and $\mathrm{SNR_{Rx}}$, \eqref{Eq:EqSR1} can be expressed in terms of $\rho$ and $\mathrm{SNR_{Rx}}$ as
\begin{align}
        S = \log_2\left(\frac{1+\mathrm{SNR_{Rx}}^{-1}}{\rho+\mathrm{SNR_{Rx}}^{-1}}\right).
\label{Eq:EqSR2}
\end{align}
For $\rho= 0$, the secrecy rate acquires its maximum value $S_\mathrm{max}\equiv \log_2(1+\mathrm{SNR_{Rx}})$, and for very large $\mathrm{SNR_{Rx}}$, $S\approx-\log_2 \rho$, i.e. depends entirely on the power ratio. Therefore, knowledge of the power ratio $\rho$ provides useful information about the expected PLS performance. Overall, the higher the SNR at the Rx and the lower the power ratio, the higher the secrecy rate.
%
%
%
\begin{figure*}[!t]
    \centering
    \includegraphics[width=\linewidth]{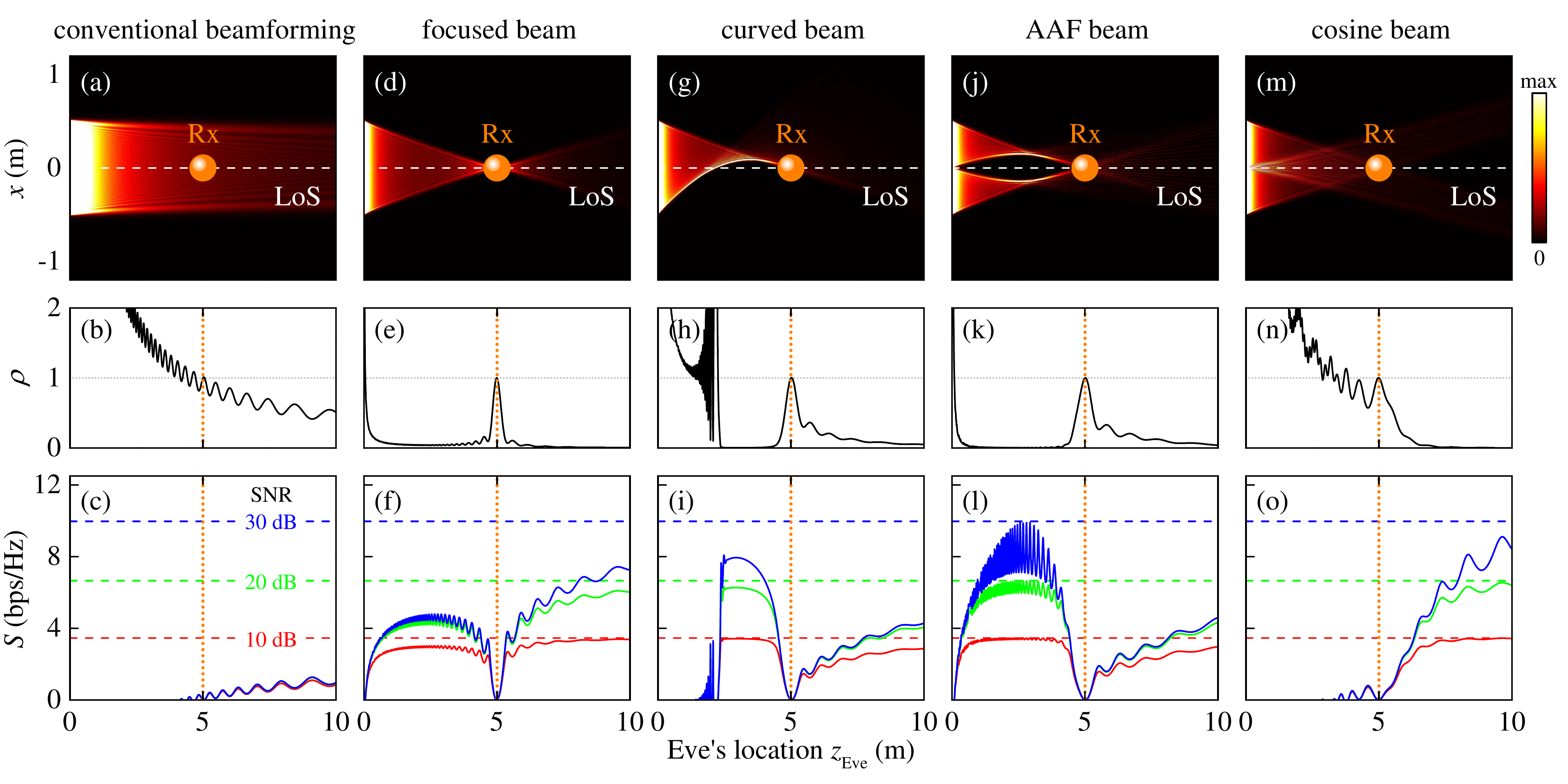}
    \caption{Near-field wavefront engineering and PLS performance for the LoS eavesdropping scenario of Fig.\,\ref{fig:fig01}. (a),(d),(g),(j),(m) Beam propagation. (b),(e),(h),(k),(n) Power ratio $\rho$, as a function of Eve's location along the Tx-Rx LoS. (c),(f),(i),(l),(o) Secrecy rate, as a function of Eve's location along the Tx-Rx LoS, for different SNR levels. The dashed lines mark $S_\mathrm{max}$ for each SNR.}
    \label{fig:fig02}
\end{figure*}
%
%
\section{PLS performance assessment}
\label{sec:PerformanceAssessment}
\noindent In this section, we assess the performance of the various beam profiles generated by the Tx, for the LoS and VoU eavesdropping scenarios presented in Fig.\,\ref{fig:fig01}. The Tx is equipped with a ULA consisting of $N_x=1000$ elements and operates at $150\,\mathrm{GHz}$ with $d_x=\lambda/2=1\,\mathrm{mm}$, i.e. has size $L_x=1\,\mathrm{m}$. The Fraunhofer distance for this ULA is $z_F=2L_x^2/\lambda\approx1\,\mathrm{km}$, which is much larger than the area where the Rx and Eve reside. Operation this deep in the near-field is ideal for efficient wavefront engineering \cite{Droulias2026}.

\subsection{LoS eavesdropping}
\noindent We begin by examining the PLS performance of the various beams, for LoS eavesdropping, i.e. when Eve moves along the Tx-Rx LoS. %
\subsubsection{Conventional beamforming}
For conventional beamforming, the beam undergoes the well-known rapid spatial intensity oscillations, as demonstrated in Fig.\,\ref{fig:fig02}(a), where \eqref{Eq:EqPHASE1} is used with $\theta_r=0^\circ$. The beam is still very deep in the Fresnel range, hence the typical far-field features have not been developed yet. For Eve on the LoS, the power ratio is presented in Fig.\,\ref{fig:fig02}(b), as a function of Eve's location. The ratio exceeds unity, and this is no surprise, because the beam is stronger closer to the Tx. In Fig.\,\ref{fig:fig02}(c) we present the corresponding secrecy rate $S$, for three SNR levels. The dashed lines mark $S_\mathrm{max}$, the maximum $S$ for each noise level. As expected, the performance is extremely poor. The secrecy rate in \eqref{Eq:EqSR2} is dominated by the very high values of $\rho$, which is why $S$ remains practically insensitive to the noise level.
\subsubsection{Focused beam}
Fig.\,\ref{fig:fig02}(d) presents the propagation of a focused beam, for which \eqref{Eq:EqPHASE2} is used with $\theta_r=0^\circ$ and $d_f=5\,\mathrm{m}\,(=z_\mathrm{Rx})$. The ratio $\rho$ presented in Fig.\,\ref{fig:fig02}(e) reveals that the power at Eve can be significantly suppressed, due to the extreme concentration of all emitted power towards the focal point, where Rx is located. This conclusion is reflected in the calculated $S$, which is presented in Fig.\,\ref{fig:fig02}(f). For low SNR, the performance reaches practically its maximum for most Eve's locations.
\subsubsection{Curved beam}
Fig.\,\ref{fig:fig02}(g) presents the propagation of a curved beam, for which the trajectory has been designed to cross the ULA at $x_\mathrm{0C}=-0.5\,\mathrm{m}\,(=-L_x/2)$, while passing through the point with $(x_P,z_P)=(0, 2\,\mathrm{m})$ to reach the Rx. Using \eqref{Eq:EqX02},\eqref{Eq:EqZ02},\eqref{Eq:EqB2}, these conditions lead to $\beta=-0.05\,\mathrm{m}^{-1}$, $x_0=0.1125\,\mathrm{m}$ and $z_0=3.5\,\mathrm{m}$, which are the beam parameters used in \eqref{Eq:EqPHASE3}. To account for the displacement of the main lobe from the trajectory, $x_0+\delta x_m$ has been used, with $\delta x_m=-0.0325\,\mathrm{m}$. As shown in Fig.\,\ref{fig:fig02}(h), in the protected region that extends between $2$--$5\,\mathrm{m}$, the curved beam is capable of offering even higher eavesdropping suppression than the focused beam. This is also identified in $S$, which is presented in Fig.\,\ref{fig:fig02}(i).
\subsubsection{AAF beam}
The AAF beam is formed using \eqref{Eq:EqPHASE3} and \eqref{Eq:EqPHASE4}, with \eqref{Eq:EqAAFx0} and \eqref{Eq:EqAAFz0}. The choice $\beta=\pm0.03\,\mathrm{m}^{-1}$ leads to $x_0=\mp 0.1875\,\mathrm{m}$ and $z_0=2.5\,\mathrm{m}$, which are the beam parameters used to propagate the beam shown in Fig.\,\ref{fig:fig02}(j). To account for the displacement of the main lobe of each individual curved beam from its respective trajectory, $x_0+\delta x_m$ has been used, with $\delta x_m=\pm 0.0482\,\mathrm{m}$. The resulting AAF beam is capable of delivering the optimum performance for Eve residing in the Tx-Rx LoS, as identified in the power ratio presented in Fig.\,\ref{fig:fig02}(k), as well as the corresponding secrecy rate presented in Fig.\,\ref{fig:fig02}(l).
%
%
%
\begin{figure*}[!b]
    \centering
    \includegraphics[width=\linewidth]{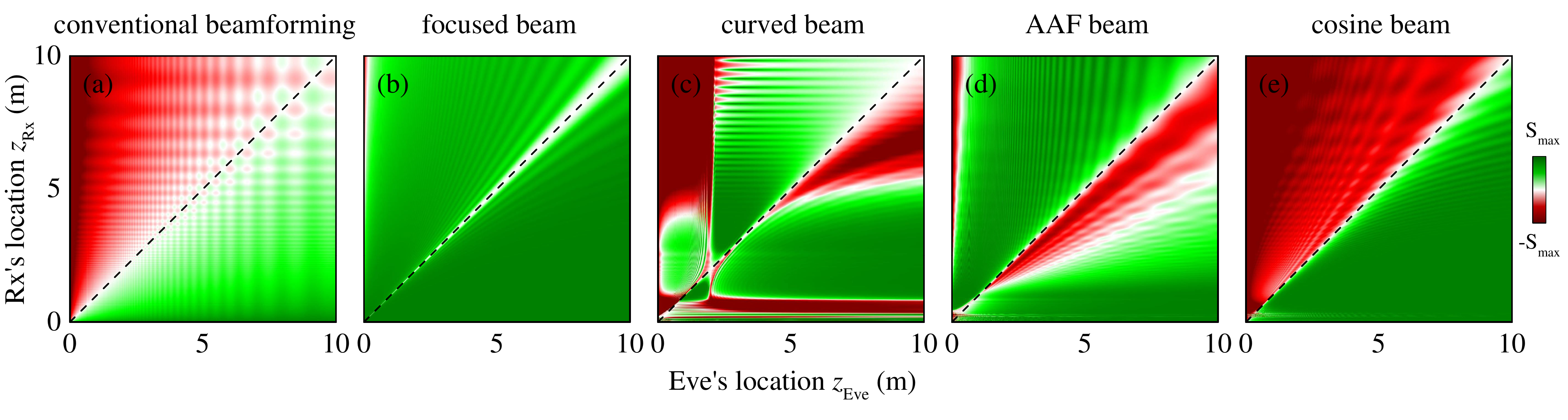}
    \caption{Secrecy rate for the LoS eavesdropping scenario of Fig.\,\ref{fig:fig01}, as a function of Eve's location along the Tx-Rx LoS, $z_\mathrm{Eve}$, for various Rx locations $z_\mathrm{Rx}$. The dashed line traces the locations where $z_\mathrm{Eve}=z_\mathrm{Rx}$, at which $\rho=1$ and, hence, $S=0$. In this example, $\mathrm{SNR=10\,\,dB}$.}
    \label{fig:fig03}
\end{figure*}
%
%
\subsubsection{Cosine beam}
Fig.\,\ref{fig:fig02}(m) presents the propagation of a cosine beam with $\theta_r=0^\circ$ and $z_\mathrm{max}=z_\mathrm{Rx}+\delta z_m$, with $\delta z_m=1.1\,\mathrm{m}$, to align the outmost maximum of the beam with the Rx. The two counter-propagating waves that constitute the beam diverge after the Rx, leaving a low power region right behind it. This can be identified in $\rho$, which is presented in Fig.\,\ref{fig:fig02}(n), and the respective $S$, which is presented in Fig.\,\ref{fig:fig02}(o). Hence, cosine beams offer the best performance when Eve is located on the Tx-Rx LoS, behind the Rx. \\
\indent The study of the various beams for a fixed Rx location naturally raises the question of how their performance depends on the exact Rx location. In Fig.\,\ref{fig:fig03} we gradually change the Rx distance from the Tx in finite steps, and for each step we calculate $S$ along the Tx-Rx LoS, as previously in the bottom panel row of Fig.\,\ref{fig:fig02}. We choose $\mathrm{SNR}=10\,\mathrm{dB}$, for which $S_\mathrm{max}\approx3.46\,\mathrm{bps/Hz}$. The secure regions $(S>0)$ are shown in green, while the vulnerable regions $(S<0)$ are shown in red. The dashed line marks the location of all points with $z_\mathrm{Rx}=z_\mathrm{Eve}$, for which $\rho=1$ and consequently $S=0$. 
Fig.\,\ref{fig:fig03}(a) presents the performance for \textit{conventional beamforming}. Regardless of the exact Rx location, the Tx-Rx region is always vulnerable, while behind the Rx the performance gives only slightly positive $S$.
The \textit{focused beam}, on the other hand, offers a wide range of secure areas, as shown in Fig.\,\ref{fig:fig03}(b). This is ensured by the fact that the focal distance is always aiming at the Rx, i.e. $d_f=z_\mathrm{Rx}$. 
The curved beam shown in Fig.\,\ref{fig:fig03}(c) is designed to pass through the point with $(x_P,z_P)=(0, 2\,\mathrm{m})$ and the Rx, as the Rx location changes. It can also deliver high performance, however within a limited range of Rx locations. This is a direct consequence of the formation efficiency of curved beams, which require large panels to fulfill sharp bends at large distances. 
The \textit{AAF beam}, shown in Fig.\,\ref{fig:fig03}(d), is designed to always focus on the Rx. It performs similarly to the focused beam, however with increasing Rx distance it fails to provide security right behind the Rx. This results from stray radiation of the two individual curved beams that constitute the AAF beam. 
Last, the \textit{cosine beam} offers an extended protected region behind the Rx, as shown in Fig.\,\ref{fig:fig03}(e). This is achieved by constantly tuning the cosine beam range to always align with the Rx location.
%
%
%
%
\begin{figure*}[!t]
    \includegraphics[width=\linewidth]{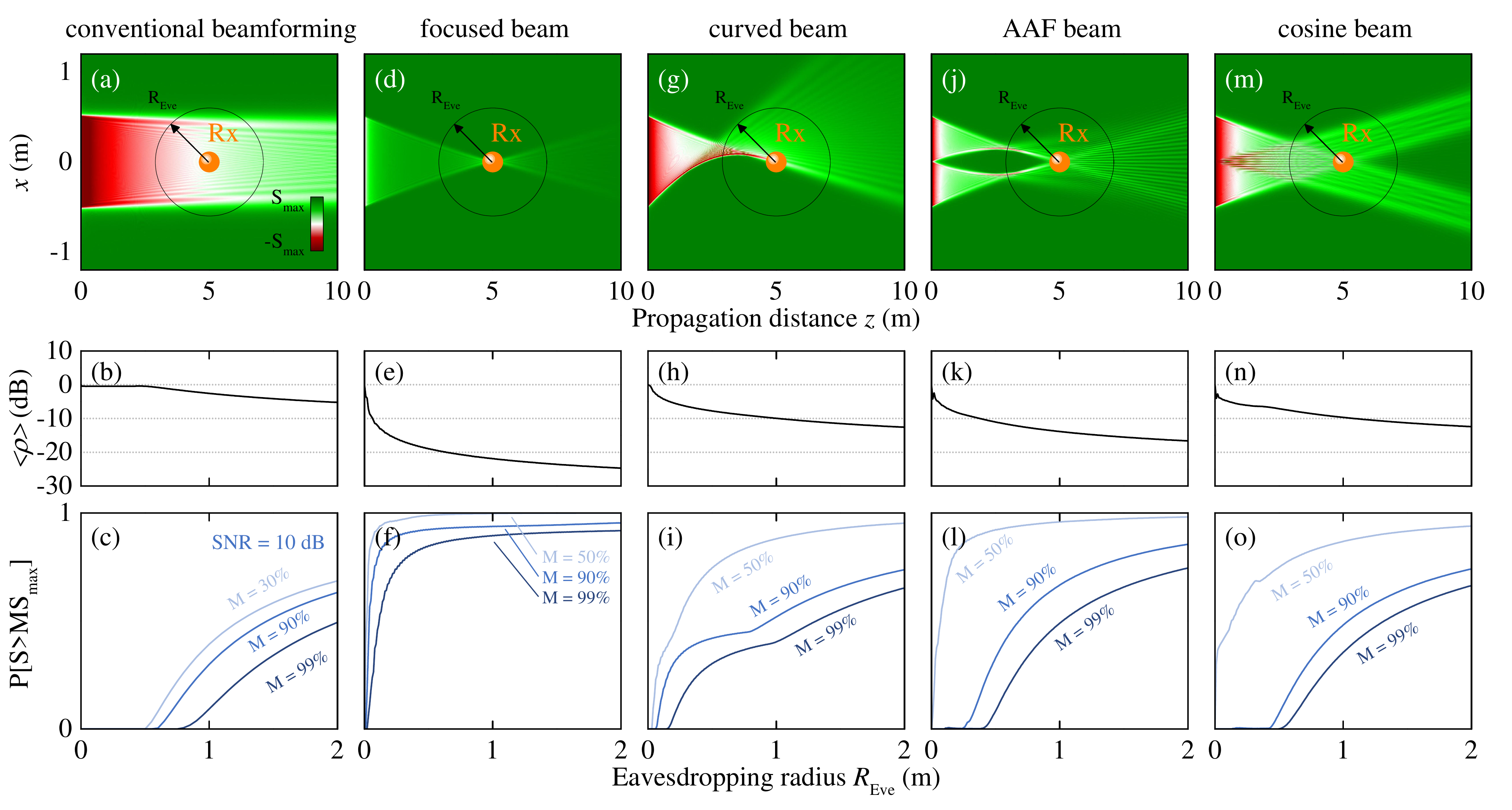}
    \caption{PLS performance for the VoU eavesdropping scenario of Fig.\,\ref{fig:fig01}. (a),(d),(g),(j),(m) Spatial map of secrecy rate for $\mathrm{SNR=10\,\,dB}$, calculated with respect to the chosen Rx location for all possible Eve locations in the shown area. (b),(e),(h),(k),(n) Power ratio $\rho$, averaged within a disk of radius $R_\mathrm{Eve}$, centered at the Rx location. (c),(f),(i),(l),(o) Probability of $S>MS_\mathrm{max}$ for different values of $M$, as a function of the disk radius within which Eve is located.}
    \label{fig:fig04}
\end{figure*}

\subsection{VoU eavesdropping}
\noindent Next, the performance of the various beams is evaluated in the VoU eavesdropping scenario of Fig.\,\ref{fig:fig01}. Figs.\,\ref{fig:fig04} (a),(d),(g),(j),(m) depict the calculated secrecy rate for all possible Eve's locations within a large disk that covers the entire shown area. Because the spatial distribution of $S$ is determined by the beam's spatial power distribution relative to its power at the Rx location, $S$ essentially imprints the beam shape in space, providing a map of secure (green) and vulnerable (red) areas. In this example, $\mathrm{SNR}=10\,\mathrm{dB}$, for which $S_\mathrm{max}\approx3.46\,\mathrm{bps/Hz}$. The circle outlines the region within which Eve is located, with uniform probability.
In Figs.\,\ref{fig:fig04} (b),(e),(h),(k),(n), the power ratio $\rho$ is averaged within the disk, and the result is presented as a function of the radius $R_\mathrm{Eve}$. With increasing $R_\mathrm{Eve}$, $\langle\rho\rangle$ reduces. This is expected due to spatial averaging over an increasingly larger area that includes progressively more low-power regions outside the beam. While all beam types exhibit the same trend, their absolute performance differs, with the focused beams delivering the highest performance, and conventional beamforming yielding the lowest. The superior performance of focused beams can be attributed to their strong spatial power concentration at the focal point, combined with the rapid power decay away from the focus. This is a unique feature that maximizes the averaged power ratio over the other beam types. 
To assess the PLS performance of the various beams, Figs.\,\ref{fig:fig04} (c),(f),(i),(l),(o) show the probability that the ratio $S/S_\mathrm{max}$ exceeds a given threshold $M$, as a function of the disk radius $R_\mathrm{Eve}$ (examples shown for $M=50\%,90\%,99\%$). For each value of $R_\mathrm{Eve}$, this probability is calculated by randomly sampling Eve's location within the disk and determining the fraction of sampled locations for which $S>MS_\mathrm{max}$. These results verify that focused beams outperform all beams in VoU eavesdropping, in accord with the conclusions drawn from the observed performance in terms of the respective $\langle\rho\rangle$.
\section{From ULAs to UPAs}
\label{sec:FromULAStoUPAS}

\noindent Compared to ULAs, UPAs offer an additional degree of freedom by extending the array along the $y$-direction. Hence, the input profile \eqref{Eq:EqINPUTWAVE} of the various beams can be applied in a two-dimensional (2D) array, to control their spatial distribution along both transverse directions.
Note, however, that, with the exception of focused beams, the phase distributions $\phi$ of the various beam types studied in this work are invariant along the $y$-direction. As a result, beam formation using UPAs yields a beam profile that is approximately uniform along the $y$-direction, and hence, PLS performance that is similar to ULAs (see Appendix for details). 
To fully exploit the additional spatial dimension offered by UPAs, independent phase distributions can be imposed along the $x$- and $y$-axes and subsequently combined to form a 2D phase distribution across the antenna aperture. This 2D phase control provides greater flexibility in shaping the radiated wavefront and controlling its spatial evolution. In essence, the phase can be decomposed as
\begin{align}
    \phi(x,y)=\phi_x(x) + \phi_y(y),
\end{align}
where $\phi_x(x)$ is any of \eqref{Eq:EqPHASE1}, \eqref{Eq:EqPHASE3}, \eqref{Eq:EqPHASE4}, \eqref{Eq:EqPHASE5}, and $\phi_y(y)$ can be chosen independently. \\
%
%
\begin{figure*}[!t]
    \centering
    \includegraphics[width=\linewidth]{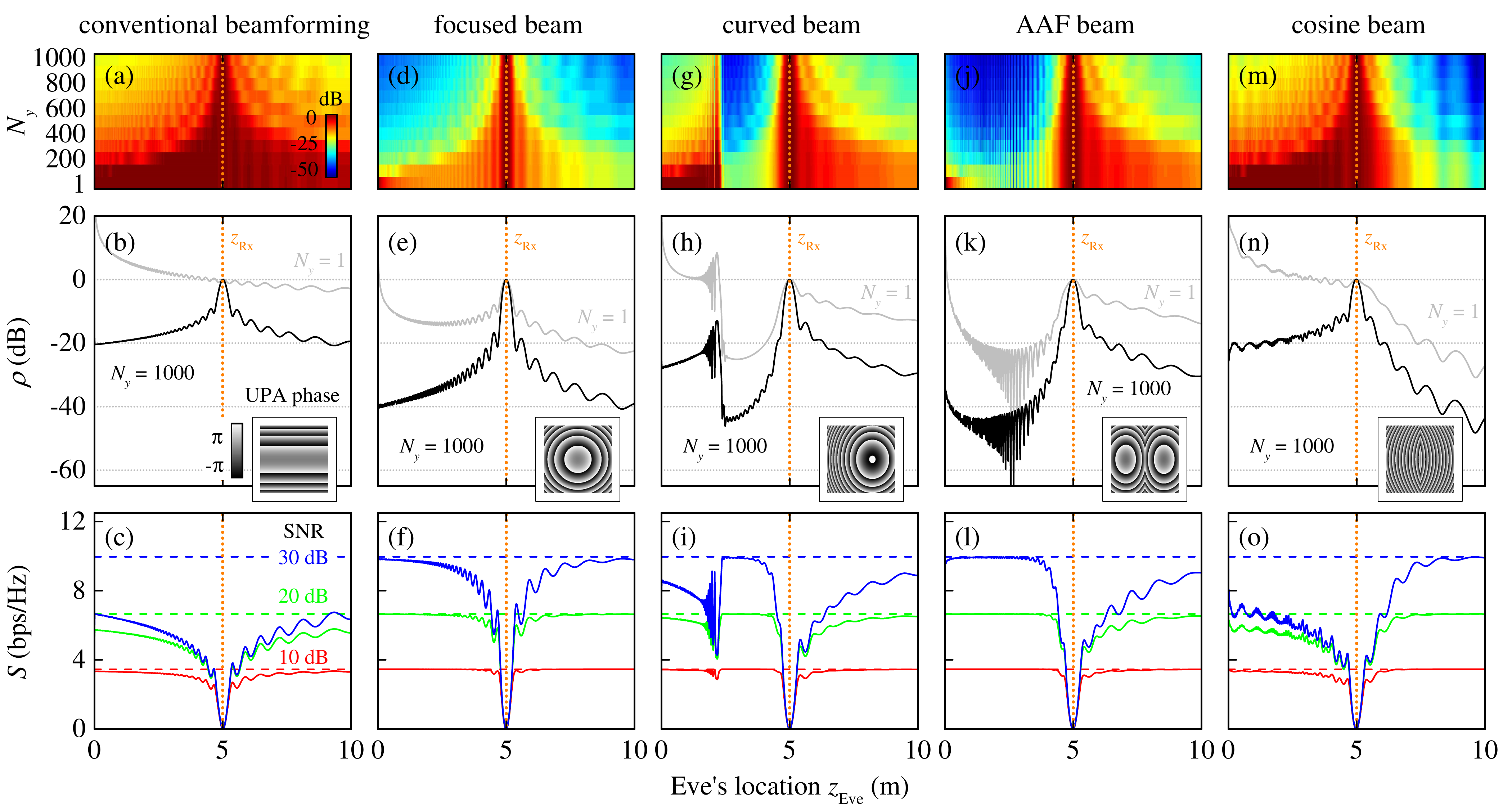}
    \caption{PLS performance for the LoS eavesdropping scenario of Fig.\,\ref{fig:fig01}, using UPAs with cylindrical focusing along the $y$-direction. (a),(d),(g),(j),(m) Power ratio $\rho$, as a function of the number of elements $N_y$ along the $y$-axis of the UPA, with $N_x=1000$. (b),(e),(h),(k),(n) Cross-sections for $N_y=1$ (ULA limit) and $N_y=1000$ (square UPA limit). The insets depict the UPA phase, when the entire panel is used. (c),(f),(i),(l),(o) Secrecy rate for $N_y=1000$, as a function of Eve's location along the Tx-Rx LoS, for different SNR levels. The dashed lines mark $S_\mathrm{max}$ for each SNR.}
    \label{fig:fig05}
\end{figure*}
%
\indent Note that, for focused beams, \eqref{Eq:EqPHASE2} already has a 2D form, which can also be decomposed to independent $x$ and $y$ components. Expanding \eqref{Eq:EqPHASE2} around the center of the UPA, i.e., at $x = 0, y = 0$, and keeping all terms up to $2^{nd}$ order leads to
\begin{align}
    \phi(x,y)=k d_f-kx\sin\theta_r + k\frac{x^2\cos^2\theta_r+y^2}{2d_f},
    \label{Eq:EqPHASE2approx}
\end{align}
which can be decomposed into 
\begin{align}
    \label{Eq:EqPHASE2approxX}
    \phi_x(x)&=-kx\sin\theta_r + k\frac{x^2\cos^2\theta_r}{2d_f}, 
    \\
    \label{Eq:EqPHASE2approxY}
    \phi_y(y)&= k\frac{y^2}{2d_f},
\end{align}
and we have omitted the global phase term $k d_f$, which multiplies the entire field by the same complex factor everywhere.
The parabolic phase \eqref{Eq:EqPHASE2approx} is a valid approximation of the exact spherical phase profile \eqref{Eq:EqPHASE2} for a wide range of practical cases \cite{Droulias2024,Droulias2025_foc}.

\subsection{Vertical focusing}

\noindent Motivated by the need to enhance the received power at the Rx while suppressing that at Eve, the additional spatial degree of freedom offered by a UPA can be exploited to further improve this power discrimination. Specifically, the phase distribution along the $x$-axis is designed to preserve and exploit the distinctive propagation characteristics of each considered beam type, while an additional focusing mechanism is introduced along the $y$-axis. The resulting cylindrical focusing concentrates the transmitted power around the Rx while reducing the power incident on Eve, thereby enhancing the power contrast between the legitimate receiver and the eavesdropper and improving the overall PLS performance. In essence, the phase $\phi_y(y)$ that imposes beam focusing in the $yz$-plane is expressed as
\begin{align}    
    \phi_y(y) = k \sqrt{y^2+ y_f^2}\approx k d_f+k\frac{y^2}{2y_f},
    \label{Eq:EqPHASEfocusY}
\end{align}
where $y_f$ is the focal distance in the $yz$-plane. $\phi_x(x)$ controls beam formation in the $xz$-plane, i.e. \eqref{Eq:EqPHASE1} for conventional beamforming, \eqref{Eq:EqPHASE3} for curved beam, \eqref{Eq:EqPHASE4} for AAF beam, and \eqref{Eq:EqPHASE5} for cosine beam. For beam focusing on the $xz$-plane, either \eqref{Eq:EqPHASE2} or the combination of \eqref{Eq:EqPHASE2approxX} and \eqref{Eq:EqPHASE2approxY} essentially lead to spherical focusing, i.e. radially symmetric phase profile. \\
%
%
%
\begin{figure}[!t]
    \centering
    \includegraphics[width=\linewidth]{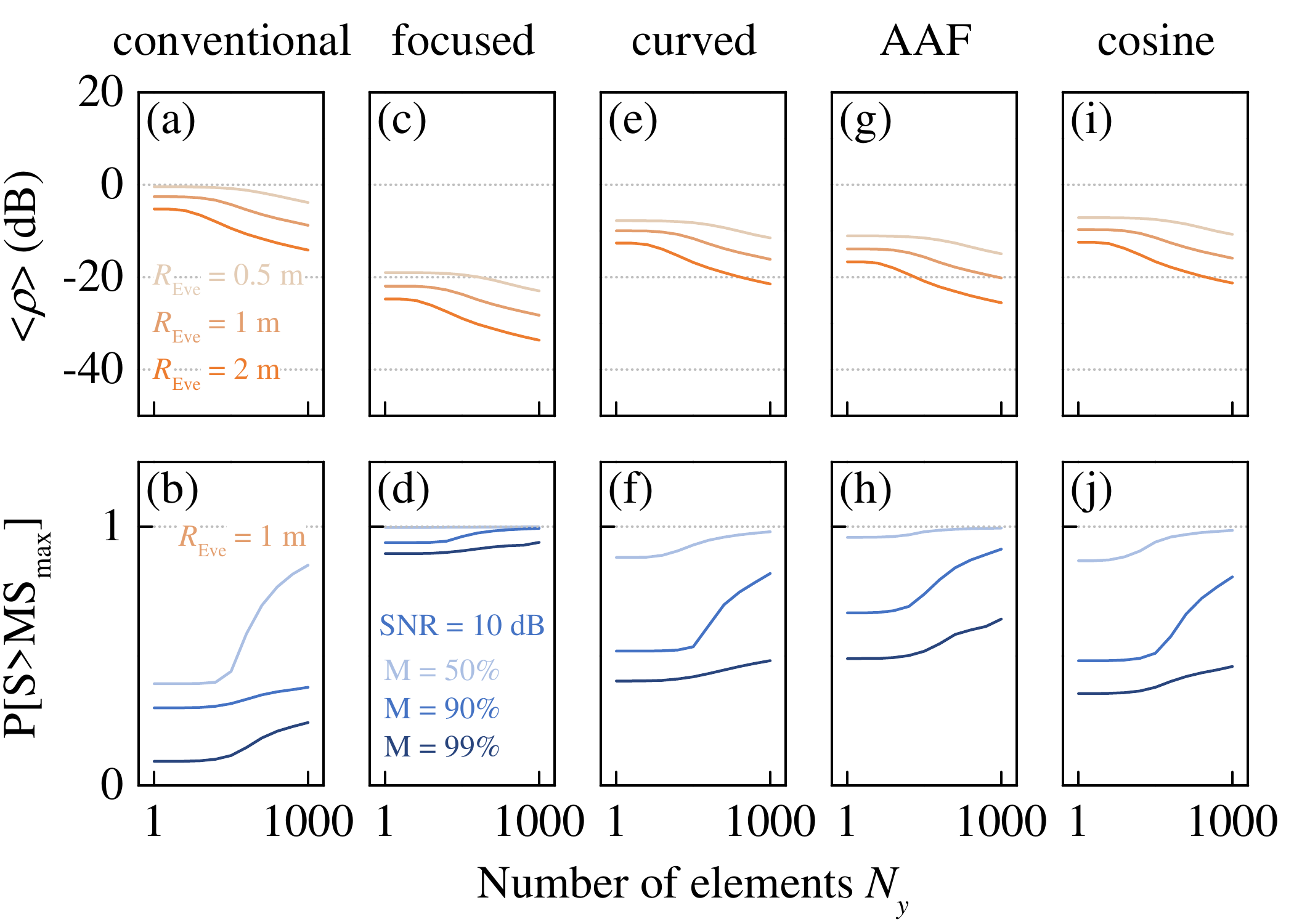}
    \caption{PLS performance for the VoU eavesdropping scenario of Fig.\,\ref{fig:fig01}, as a function of the number of elements $N_y$ along the $y$-axis of the UPA, leveraging cylindrical focusing along the $y$-direction. (a),(c),(e),(g),(i) Power ratio $\rho$, averaged within a disk of radius $R_\mathrm{Eve}$, centered at the Rx location. (b),(d),(f),(h),(j) Probability of $S>MS_\mathrm{max}$ for different values of $M$, for $R_\mathrm{Eve}=1\,\mathrm{m}$ and $\mathrm{SNR}=10\,\mathrm{dB}$.}

    \label{fig:fig06}
\end{figure}
\indent In Fig.\,\ref{fig:fig05} we study the various beam types using vertical (cylindrical) focusing for the LoS scenario. For all beams, we set $y_f=z_\mathrm{Rx}$, in order to focus the beam onto the Rx. In Figs.\,\ref{fig:fig05}(a),(d),(g),(j),(m) we calculate $\rho$ as a function of the number of elements $N_y$ along the $y$-axis of the UPA, while keeping $N_x=1000$. Clearly, with increasing UPA height, cylindrical focusing gradually suppresses the power ratio $\rho$ in the vicinity of the Rx, for all beam types. Explicit cross-sections for $N_y=1$ (ULA limit) and $N_y=1000$ are presented in Figs.\,\ref{fig:fig05}(b),(e),(h),(k),(n); the insets depict the UPA phase, when the entire panel is used. 
This improvement is also reflected in the secrecy rates, which are evaluated for $N_y=1000$ as a function of Eve's position along the Tx--Rx LoS for different SNR levels, and are shown in Figs.\,\ref{fig:fig05}(c),(f),(i),(l),(o). The focused beams consistently achieve the highest secrecy rates, while all considered beam types exhibit a substantial improvement over their corresponding ULA-based counterparts. \\
\indent In Fig.\,\ref{fig:fig06} we evaluate the performance of these beams in the VoU scenario. In Fig.\,\ref{fig:fig06}(a),(c),(e),(g),(i) we present $\langle\rho\rangle$, as a function of $N_y$, with $N_x=1000$, calculated for selected disk radii. In Fig.\,\ref{fig:fig06}(b),(d),(f),(h),(j) we select $R_\mathrm{Eve}=1\,\mathrm{m}$, and we present the probability $P[S>MS_\mathrm{max}]$ for $\mathrm{SNR}=10\,\mathrm{dB}$. Clearly, with increasing UPA height, cylindrical focusing suppresses the averaged power ratio $\langle\rho\rangle$ in the vicinity of the Rx, for all beam types, in turn improving $S$. The focused beams consistently achieve the best performance, with all considered beam types benefiting substantially from the transition from ULA- to UPA-based implementations.

\subsection{Hollow-focused beams}
\begin{figure}[!t]
    \centering
    \includegraphics[width=\linewidth]{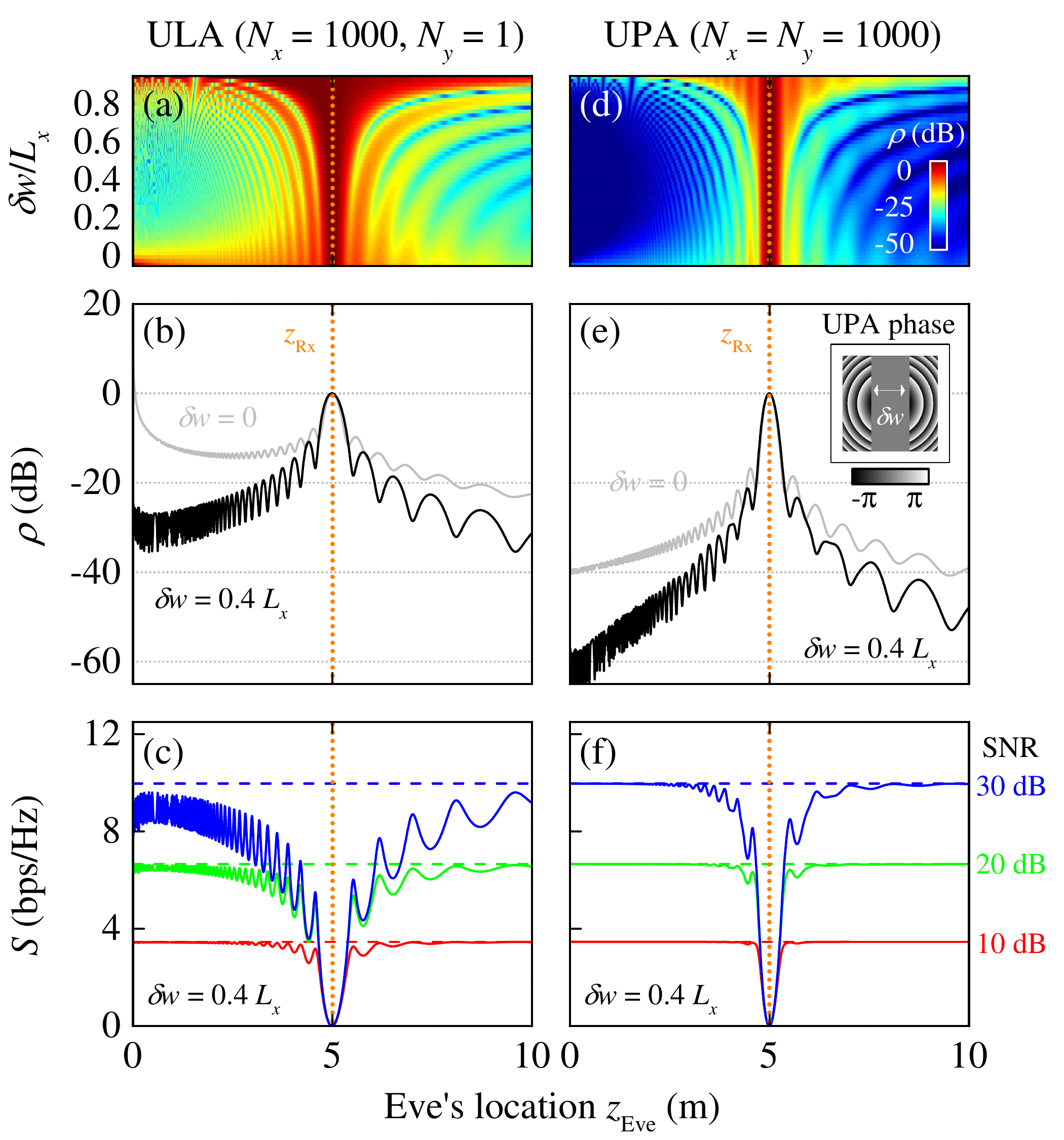}
    \caption{Focused beams with a central inactive strip, and PLS performance for the LoS eavesdropping scenario of Fig.\,\ref{fig:fig01}. (a),(d) Power ratio $\rho$, as a function of the normalized inactive strip width $\delta w/L_x$. (b),(e) Cross-sections for $\delta w=0$ (conventional focusing) and $\delta w=0.4L_x$ (optimum strip width). The inset in (e) depicts the UPA phase. (c),(f) Secrecy rate for the optimum strip width, as a function of Eve's location along the Tx-Rx LoS, for different SNR levels. The dashed lines mark $S_\mathrm{max}$ for each SNR.}
    \label{fig:fig07}
\end{figure}

\begin{figure}[!b]
    \centering
    \includegraphics[width=\linewidth]{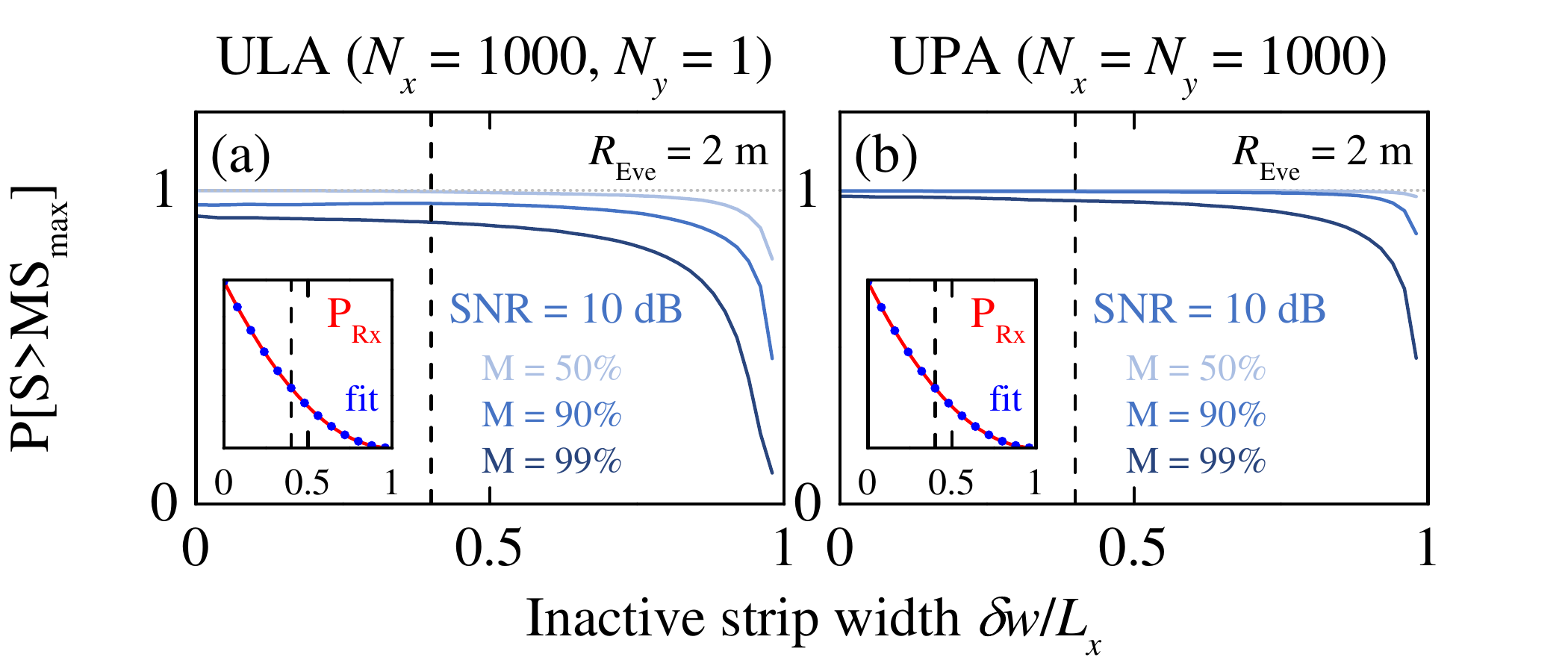}
    \caption{PLS performance for the VoU eavesdropping scenario of Fig.\,\ref{fig:fig01}, using focused beams with a central inactive strip. Probability of $S>MS_\mathrm{max}$, as a function of $\delta w/L_x$, for (a) ULA with $N_x=1000,N_y=1$ and (b) UPA with $N_x=N_y=1000$. In this example, $R_\mathrm{Eve}=2\,\mathrm{m}$ and $\mathrm{SNR}=10\,\mathrm{dB}$. \textit{Inset}: Numerically calculated power at the Rx (solid red line) and analytical fit (blue dots), as a function of $\delta w/L_x$.}
    \label{fig:fig08}
\end{figure}

\noindent The low-power region exhibited by AAF beams between the Tx and Rx provides spatial protection against LoS eavesdropping, as the transmitted power remains suppressed over the region separating the legitimate terminals. 
Inspired by this property, a similar protected region can be introduced for focused beams through a central inactive strip configuration. Specifically, the UPA is operated using all elements except those contained within a narrow strip centered on the $x$-axis and having a width $\delta w$ along the $x$-direction, i.e.,
\begin{align}    
    A(x) = 0, \,\,\,|x|<\frac{\delta w}{2}
    \label{Eq:EqPHASEhollowX}
\end{align}
\noindent For a relatively narrow inactive strip, the resulting aperture is expected to preserve the focusing capability of the beam, while creating a low-power region along the Tx--Rx path, thereby combining the power-concentration benefits of focused beams with the spatial protection offered by a hollow transmission region. The formation of the low-power region between the Tx and Rx makes this configuration particularly suited to LoS eavesdropping and can be applied to both ULAs and UPAs. \\
\indent To evaluate the performance of hollow-focused beams in the LoS scenario, in Fig.\,\ref{fig:fig07}(a), and Fig.\,\ref{fig:fig07}(d) we calculate $\rho$ as a function of the inactive strip width, $\delta w$, for the ULA ($N_x=1000,N_y=1$) and the UPA ($N_x=N_y=1000$), respectively. Increasing $\delta w$ surely extends the width of the protected region along the $x$-direction, however, increases the number of inactive array elements, therefore reduces the focusing efficiency and, in turn, the power focused at the Rx; in the limit $\delta w=L_x$, all array elements are deactivated, resulting in no beam formation. Due to this trade-off, the calculated power ratio reaches an optimum roughly at $\delta w\approx 40\%L_x$. Figs.\,\ref{fig:fig07}(b,e) show explicit cross-sections of Figs.\,\ref{fig:fig07}(a,d), respectively, at this optimum value; the inset depicts the UPA phase, also showing the inactive strip configuration. This improvement is also reflected in the secrecy rates shown in Figs.\,\ref{fig:fig07}(c,f), evaluated as a function of Eve's position along the Tx--Rx LoS for different SNR levels. A comparison of Fig.\,\ref{fig:fig07}(c) with Fig.\,\ref{fig:fig02}(f) for the ULA and Fig.\,\ref{fig:fig07}(f) with Fig.\,\ref{fig:fig05}(f) for the UPA, verifies that hollow-strip focusing further boosts the LoS PLS performance. \\
\indent In Fig.\,\ref{fig:fig08}(a) and Fig.\,\ref{fig:fig08}(b) we revisit the VoU scenario for the hollow-strip focused beams, using the ULA with $N_x=1000,N_y=1$ and the UPA with $N_x=N_y=1000$, respectively. We choose $R_\mathrm{Eve}=2\,\mathrm{m}$, and we calculate the probability $P[S>MS_\mathrm{max}]$, as a function of the inactive strip width normalized to the array width, $\delta w/L_x$. In contrast to the LoS scenario, no similar trade-off is observed as $\delta w$ changes, between the width of the protected region and the focusing efficiency. Instead, the PLS performance decreases monotonically. This behavior can be attributed to the fact that, while the width of the protected region increases linearly with $\delta w$, the focusing efficiency decreases nonlinearly, as evidenced by the insets showing the received power at the Rx, $P_\mathrm{Rx}$. The focused power follows a parabolic decay
\begin{align}    
    P_\mathrm{Rx}\propto (1-\frac{\delta w}{L_x})^2, \,\,\,0<\frac{\delta w}{L_x}<1
    \label{Eq:EqPRx}
\end{align}
as confirmed by the fit shown by the blue dots. Consequently, for small inactive regions, where $P_\mathrm{Rx}$ is approximated as $P_\mathrm{Rx}\approx 1-2(\delta w/L_x)$, the received power at the Rx decreases linearly with $\delta w$, thereby compensating for the linear increase in the width of the protected region and resulting in nearly unchanged PLS performance. At the optimum width, $\delta w/L_x=0.4$, the PLS performance remains comparable to that of the fully focused beam. Hence, hollow-focused beams that are particularly advantageous in LoS scenarios, can also provide effective performance in VoU scenarios.
%
%
%
%
%
%


\begin{table}[t]
\centering
\caption{Relative PLS performance of beams studied in this work}
\label{tab1}

\begin{threeparttable}

\begin{tabular}{c c c c c}
\hline
Array type & Beam type &
\multicolumn{3}{c}{Eavesdropping scenario} \\
\cline{3-5}
  &   &
\multicolumn{2}{c}{LoS} & VoU \\
\cline{3-4}
& &
$z_{\mathrm{Eve}} < z_{\mathrm{Rx}}$ &
$z_{\mathrm{Eve}} > z_{\mathrm{Rx}}$ &
\\
\hline

\multirow{6}{*}{ULA}
& Conventional
& \inferior
& \inferior
& \inferior \\

& Focused
& \intermediate
& \superior
& \superior \\

& Curved
& \superior
& \intermediate
& \intermediate \\

& AAF
& \superior
& \intermediate
& \intermediate \\

& Cosine
& \inferior
& \superior
& \intermediate \\

& Hollow-focused
& \superior
& \superior
& \superior \\

\hline

\multirow[c]{6}{*}{%
    \parbox{1.75cm}{\centering UPA with\\ vertical \\ focusing}%
}
& Conventional
& \intermediate
& \intermediate
& \inferior \\

& Focused
& \superior
& \superior
& \superior \\

& Curved
& \superior
& \superior
& \intermediate \\

& AAF
& \superior
& \superior
& \intermediate \\

& Cosine
& \intermediate
& \superior
& \intermediate \\

& Hollow-focused
& \superior
& \superior
& \superior \\

\hline

\end{tabular}

\begin{tablenotes}
\footnotesize
\centering
\item $\checkmark$: Highly suitable;\quad
$\circ$: Moderately suitable;\quad
$\times$: Less suitable.
\end{tablenotes}

\end{threeparttable}
\end{table}

\section{Discussion and conclusion}
\label{sec:Conclusion}
Wavefront engineering can substantially enhance PLS performance in near-field scenarios by tailoring the beam profile in the vicinity of the eavesdropper. Conversely, the results obtained with conventional beamforming demonstrate that the absence of such wavefront control can significantly degrade PLS performance. Specifically, curved beams and AAF beams are well-suited for LoS eavesdropping when $z_\mathrm{Eve}<z_\mathrm{Rx}$, whereas cosine beams are more effective when $z_\mathrm{Eve}>z_\mathrm{Rx}$. Focused beams, which concentrate the transmitted power at the intended receiver, can outperform these beam types. Building on this property, hollow-focused beams combine the characteristics of AAF beams and focused beams, achieving the highest PLS performance among all beam types investigated in this work in both LoS and VoU eavesdropping scenarios. Furthermore, incorporating vertical focusing can substantially improve PLS performance across all beam types, including those based on conventional beamforming. Table~\ref{tab1} summarizes the PLS performance of the beam types studied in this work.

In this work we investigated the potential of near-field wavefront engineering for enhancing physical layer security in wireless communications. Several beam types, including focused, curved, abruptly autofocusing, and cosine beams, were generated using uniform planar arrays and systematically evaluated in terms of their ability to enhance the received power at the legitimate receiver while suppressing it at a potential eavesdropper. Power-based metrics were introduced to characterize the resulting security performance in both LoS and VoU eavesdropping scenarios. The results demonstrated that exploiting the additional spatial degrees of freedom of UPAs can substantially improve the performance of the considered beam types compared with their ULA-based counterparts, with focused beams generally providing the highest power discrimination. Building upon the spatially protected region offered by AAF beams, hollow-focused beams were subsequently introduced by deactivating a narrow strip of the transmitting aperture. This approach combines the power-concentration capability of focused beams with a low-power region between the transmitter and receiver, providing enhanced protection against LoS eavesdropping while maintaining competitive performance in VoU scenarios. Overall, the results demonstrate that tailored near-field wavefronts and aperture engineering provide promising degrees of freedom for improving physical layer security in future wireless systems.

%
%
\begin{figure*}
    \centering
    \includegraphics[width=\linewidth]{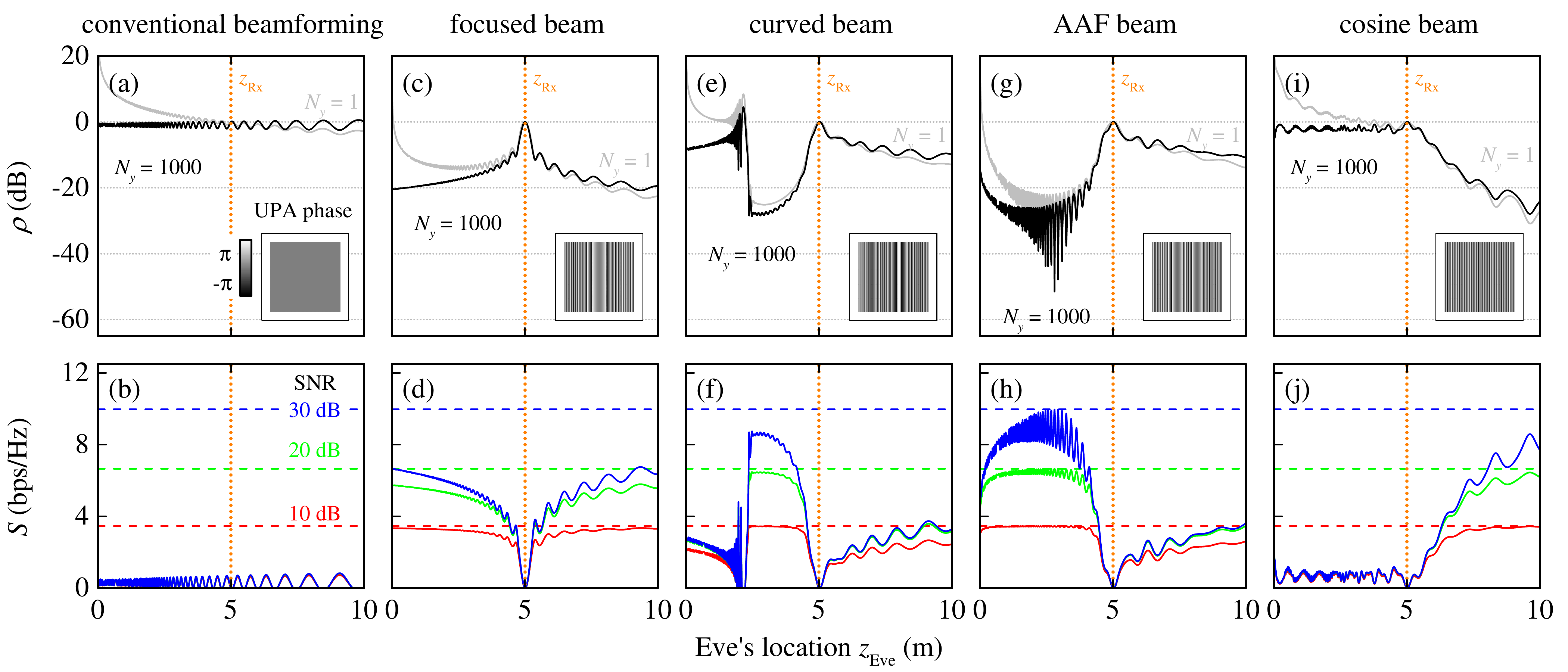}
    \caption{Near-field wavefront engineering with UPAs, and PLS performance for the LoS eavesdropping scenario of Fig.\,\ref{fig:fig01}, without application of a phase distribution along the $y$-direction. (a),(c),(e),(g),(i) Power ratio $\rho$ for a ULA with $N_x=1000,N_y=1$, and a UPA with $N_x=N_y=1000$. The insets depict the UPA phase. (b),(d),(f),(h),(j) Secrecy rate for the UPA, as a function of Eve's location along the Tx-Rx LoS, for different SNR levels. The dashed lines mark $S_\mathrm{max}$ for each SNR. Comparison with Fig.\,\ref{fig:fig02} demonstrates that the UPA does not offer any substantial benefit over ULA-based operation.}
    \label{fig:fig09}
\end{figure*}
%

\section*{Acknowledgments}
This work was supported by the European Commission’s Horizon Europe Programme under the Smart Networks and Services Joint Undertaking INSTINCT project (Grant Agreement $101139161$).

\section*{APPENDIX}

\label{Sec:AppendixA}

In Fig.\,\ref{fig:fig09} we study the various beam types for the UPA ($N_x=N_y=1000$), on which we do not impose any phase along the vertical array direction ($\phi_y(y)=0$); hence, all phases are $y$-invariant, leading to practically uniform beam shape in the vertical UPA direction. In Figs.\,\ref{fig:fig09}(a),(c),(e),(g),(i) we present the power ratio $\rho$ for the LoS scenario; the insets depict the UPA phase for each individual beam type. The respective results for the ULA have been reported in Figs.\,\ref{fig:fig05}(b),(e),(h),(k),(n), and are also shown here for easier comparison. Clearly, the transition from the ULA to the UPA does not improve $\rho$. This is also reflected in the secrecy rate, which is shown in Figs.\,\ref{fig:fig09}(b),(d),(f),(h),(j), as a function of Eve's location along the Tx-Rx LoS, for different SNR levels. Comparison with Figs.\,\ref{fig:fig02}(c),(f),(i),(l),(o) shows that, when the additional degree of freedom offered by the vertical dimension of the UPA is not exploited, the UPA does not offer any substantial benefit in PLS over ULA-based operation.

\bibliographystyle{IEEEtran}
\bibliography{IEEEabrv,main}

\begin{thebibliography}{10}
\providecommand{\url}[1]{#1}
\csname url@samestyle\endcsname
\providecommand{\newblock}{\relax}
\providecommand{\bibinfo}[2]{#2}
\providecommand{\BIBentrySTDinterwordspacing}{\spaceskip=0pt\relax}
\providecommand{\BIBentryALTinterwordstretchfactor}{4}
\providecommand{\BIBentryALTinterwordspacing}{\spaceskip=\fontdimen2\font plus
\BIBentryALTinterwordstretchfactor\fontdimen3\font minus \fontdimen4\font\relax}
\providecommand{\BIBforeignlanguage}[2]{{%
\expandafter\ifx\csname l@#1\endcsname\relax
\typeout{** WARNING: IEEEtran.bst: No hyphenation pattern has been}%
\typeout{** loaded for the language `#1'. Using the pattern for}%
\typeout{** the default language instead.}%
\else
\language=\csname l@#1\endcsname
\fi
#2}}
\providecommand{\BIBdecl}{\relax}
\BIBdecl

\bibitem{Mukherjee2014}
A.~Mukherjee, S.~A.~A. Fakoorian, J.~Huang, and A.~L. Swindlehurst, ``Principles of physical layer security in multiuser wireless networks: A survey,'' \emph{IEEE Communications Surveys \& Tutorials}, vol.~16, no.~3, pp. 1550--1573, 2014.

\bibitem{Hanzo2016}
Y.~Zou, J.~Zhu, X.~Wang, and L.~Hanzo, ``A survey on wireless security: Technical challenges, recent advances, and future trends,'' \emph{Proceedings of the IEEE}, vol. 104, no.~9, pp. 1727--1765, 2016.

\bibitem{Schaefer2017}
H.~V. Poor and R.~F. Schaefer, ``Wireless physical layer security,'' \emph{Proceedings of the National Academy of Sciences}, vol. 114, no.~1, pp. 19--26, 2017.

\bibitem{Niu2026}
H.~Niu, Y.~Xiao, X.~Lei, J.~Chen, Z.~Xiao, M.~Li, and C.~Yuen, ``A survey on artificial noise for physical layer security: Opportunities, technologies, guidelines, advances, and trends,'' \emph{IEEE Communications Surveys \& Tutorials}, vol.~28, pp. 341--381, 2026.

\bibitem{Khisti2010}
A.~Khisti and G.~W. Wornell, ``Secure transmission with multiple antennas {I}: The {MISOME} wiretap channel,'' \emph{IEEE Transactions on Information Theory}, vol.~56, no.~7, pp. 3088--3104, 2010.

\bibitem{Zurita2011}
N.~Romero-Zurita, M.~Ghogho, and D.~McLernon, ``Physical layer security of {MIMO–OFDM} systems by beamforming and artificial noise generation,'' \emph{Physical Communication}, vol.~4, no.~4, pp. 313--321, 2011, special issue on Advances in MIMO-OFDM.

\bibitem{Sharma2022}
H.~Sharma, N.~Kumar, and R.~Tekchandani, ``Physical layer security using beamforming techniques for {5G} and beyond networks: A systematic review,'' \emph{Physical Communication}, vol.~54, p. 101791, 2022.

\bibitem{Eldar2022}
H.~Zhang, N.~Shlezinger, F.~Guidi, D.~Dardari, M.~F. Imani, and Y.~C. Eldar, ``Beam focusing for near-field multiuser {MIMO} communications,'' \emph{IEEE Transactions on Wireless Communications}, vol.~21, no.~9, pp. 7476--7490, 2022.

\bibitem{Dai2023}
M.~Cui, Z.~Wu, Y.~Lu, X.~Wei, and L.~Dai, ``Near-field {MIMO} communications for {6G}: Fundamentals, challenges, potentials, and future directions,'' \emph{IEEE Communications Magazine}, vol.~61, no.~1, pp. 40--46, 2023.

\bibitem{Droulias2024}
S.~Droulias, G.~Stratidakis, and A.~Alexiou, ``Near-field engineering in {RIS}-aided links: Beamfocusing analytical performance assessment,'' \emph{IEEE Access}, vol.~12, pp. 29\,536--29\,546, 2024.

\bibitem{Guerboukha2024}
H.~Guerboukha, B.~Zhao, Z.~Fang, E.~Knightly, and D.~M. Mittleman, ``Curving {THz} wireless data links around obstacles,'' \emph{Communications Engineering}, vol.~3, no.~1, p.~58, Mar 2024.

\bibitem{Droulias2025}
S.~Droulias, G.~Stratidakis, and A.~Alexiou, ``Bending beams for {6G} near-field communications,'' \emph{IEEE Transactions on Wireless Communications}, vol.~24, no.~2, pp. 1467--1480, 2025.

\bibitem{Uchimura2026}
S.~Uchimura, J.~M. Jornet, and K.~Ishibashi, ``Optimization and characterization of near-field beams with uniform linear arrays,'' \emph{IEEE Transactions on Wireless Communications}, vol.~25, pp. 6904--6920, 2026.

\bibitem{Droulias2025_cosine}
S.~Droulias, G.~Stratidakis, and A.~Alexiou, ``Orthogonal beams and codebook design for near-field space-division multiple access in {6G} systems,'' \emph{IEEE Transactions on Antennas and Propagation}, vol.~73, no.~9, pp. 6899--6913, 2025.

\bibitem{Martínez2022}
G.~J. Anaya-López, J.~P. González-Coma, and F.~J. López-Martínez, ``Spatial degrees of freedom for physical layer security in {XL-MIMO},'' in \emph{2022 IEEE 95th Vehicular Technology Conference: (VTC2022-Spring)}, 2022, pp. 1--5.

\bibitem{Liu2023}
Y.~Liu, J.~Xu, Z.~Wang, X.~Mu, and L.~Hanzo, ``Near-field communications: What will be different?'' \emph{arXiv:2303.04003}, 2023.

\bibitem{Ferreira2023}
J.~Ferreira, J.~Guerreiro, R.~Dinis, and M.~M. Da~Silva, ``On the jamming rejection features of near-field beamforming,'' in \emph{2023 IEEE 97th Vehicular Technology Conference (VTC2023-Spring)}, 2023, pp. 1--5.

\bibitem{Ferreira2024}
J.~Ferreira, J.~Guerreiro, and R.~Dinis, ``Physical layer security with near-field beamforming,'' \emph{IEEE Access}, vol.~12, pp. 4801--4811, 2024.

\bibitem{Zhang2024}
Z.~Zhang, Y.~Liu, Z.~Wang, X.~Mu, and J.~Chen, ``Physical layer security in near-field communications,'' \emph{IEEE Transactions on Vehicular Technology}, vol.~73, no.~7, pp. 10\,761--10\,766, 2024.

\bibitem{Eldar2024}
Y.~Zhang, H.~Zhang, S.~Xiao, W.~Tang, and Y.~C. Eldar, ``Near-field wideband secure communications: An analog beamfocusing approach,'' \emph{IEEE Transactions on Signal Processing}, vol.~72, pp. 2173--2187, 2024.

\bibitem{DrouliasPLS_2024}
S.~Droulias, G.~Stratidakis, and A.~Alexiou, ``Beam-focusing in near-field communications: enhancing the physical layer security,'' in \emph{2024 IEEE 25th International Workshop on Signal Processing Advances in Wireless Communications (SPAWC)}, 2024, pp. 216--220.

\bibitem{Zhao2025}
B.~Zhao, C.~Ouyang, X.~Zhang, and Y.~Liu, ``Performance analysis of physical layer security: From far-field to near-field,'' \emph{IEEE Transactions on Wireless Communications}, vol.~24, no.~9, pp. 7392--7408, 2025.

\bibitem{Zhang2026}
Y.~Zhang, Y.~Fang, C.~You, Y.-J. Angela~Zhang, and H.~Cheung~So, ``Performance analysis and low-complexity beamforming design for near-field physical layer security,'' \emph{IEEE Transactions on Communications}, vol.~74, pp. 781--796, 2026.

\bibitem{Petrov2024}
V.~Petrov, H.~Guerboukha, D.~M. Mittleman, and A.~Singh, ``Wavefront hopping: An enabler for reliable and secure near field terahertz communications in {6G} and beyond,'' \emph{IEEE Wireless Communications}, vol.~31, no.~1, pp. 48--55, 2024.

\bibitem{Petrov2025}
V.~Petrov, H.~Guerboukha, A.~Singh, and J.~M. Jornet, ``Wavefront hopping for physical layer security in {6G} and beyond near-field {THz} communications,'' \emph{IEEE Transactions on Communications}, vol.~73, no.~5, pp. 2996--3012, 2025.

\bibitem{Stults2026}
C.~Stults, K.~K. Mutai, Z.~Fang, and D.~M. Mittleman, ``Secrecy capacity of {Airy} beams,'' \emph{IEEE Transactions on Terahertz Science and Technology}, pp. 1--7, 2026.

\bibitem{DrouliasPLS_2026}
\BIBentryALTinterwordspacing
S.~Droulias, G.~Stratidakis, and A.~Alexiou, ``Enhancing the physical layer security with bending beams,'' 2026. [Online]. Available: \url{https://arxiv.org/abs/2604.20583}
\BIBentrySTDinterwordspacing

\bibitem{Droulias2024_DM}
S.~Droulias and A.~Alexiou, ``Reconfigurable intelligent surface: {MIMO} or radiating sheet?'' \emph{IEEE Transactions on Wireless Communications}, vol.~23, no.~4, pp. 2726--2739, 2024.

\bibitem{Goel2008}
S.~Goel and R.~Negi, ``Guaranteeing secrecy using artificial noise,'' \emph{IEEE Transactions on Wireless Communications}, vol.~7, no.~6, pp. 2180--2189, 2008.

\bibitem{Choi2023}
E.~Choi, M.~Oh, J.~Choi, J.~Park, N.~Lee, and N.~Al-Dhahir, ``Joint precoding and artificial noise design for {MU-MIMO} wiretap channels,'' \emph{IEEE Transactions on Communications}, vol.~71, no.~3, pp. 1564--1578, 2023.

\bibitem{Barros2006}
J.~Barros and M.~R. D.~Rodrigues, ``Secrecy capacity of wireless channels,'' in \emph{2006 IEEE International Symposium on Information Theory}, 2006, pp. 356--360.

\bibitem{Droulias2026}
\BIBentryALTinterwordspacing
S.~Droulias and A.~Alexiou, ``Bending beams behind corners: mechanisms, challenges and capabilities for wireless connectivity,'' 2026. [Online]. Available: \url{https://arxiv.org/abs/2606.00678}
\BIBentrySTDinterwordspacing

\bibitem{Droulias2025_foc}
------, ``Orthogonal beam-focusing with planar arrays for multiple access in near-field communications,'' \emph{IEEE Open Journal of the Communications Society}, vol.~6, pp. 7319--7332, 2025.

\end{thebibliography}

\end{document}